In this paper the method for a problem solution of expenditures reduction of computing resources and time is developed at a pattern recognition. The way of construction of the minimum tests sets or separate minimum tests on Boolean matrixes is suggested. Some minimum tests can be received by using a consequence for improvement of the minimum test length from the theorem given in paper. It is possible to construct none minimum deadlock tests on the basis of the minimum tests.

-----

В этой статье разработан метод решения проблемы уменьшения затрат вычислительных ресурсов и времени при распознавании образов. Предложен метод построения множеств минимальных тестов или отдельных минимальных тестов на булевых матрицах. Отдельные минимальные тесты можно получить, если использовать для уточнения длины минимального теста следствие из теоремы, данной в статье. На основе минимальных тестов можно построить и неминимальные тупиковые тесты.

**Юлия Анатольевна Бродская**

**Формирование множеств минимальных тестов
или отдельных минимальных тестов**

### Введение

Как известно, задачи распознавания образов и диагностирования – одни из наиболее важных. Успехи, достигнутые в дискретной математике, алгебраической топологии, теории вероятностей и математической статистике, дали теоретические и практические результаты в технической диагностике, распознавании рудных месторождений, прогнозировании исходов процессов в производстве, в биологических системах. Большой вклад в теорию и практику распознавания внесли: Гренандер У., Фу К., Журавлев Ю.И., Загоруйко Н.Г., Кудрявцев В.Б., Ф.Розенблат, Богомолов А.М., Закревский А.Д., Вапник В.Н., Цыпкин Я.З., Горелик А.Л., Яблонский С.В. и др.

Под распознаванием образа объекта (объекта) понимается исследование, целью которого является определение класса объектов (предметов, явлений, процессов) $K_j \in \{K_1, K_2, ..., K_l\}$ классификации, заданной на генеральной совокупности $m$ некоторой предметной области, которому принадлежит распознаваемый объект.

Определение принадлежности распознаваемого объекта к классу объектов основывается на оценке близости его свойств (признаков) к свойствам объектов класса. Описание класса представляет собой неупорядоченный набор описаний эталонных объектов, принадлежность которых к классу известна. Описание *эталонного* объекта – строка таблицы эталонных объектов, элементы которой являются значениями признаков, описывающих объекты; множество признаков (упорядоченное), описывающее объекты – единое для всех классов. В большинстве случаев рассматриваются классификации с непересекающимися классами (разбиения): $K_{j_1} \cap K_{j_2} \neq \varnothing$. Обратим внимание на следующие особенности постановки проблемы. Разбиение генеральной совокупности $m$ определено не полностью: существуют объекты $s \in m$, принадлежность которых классам не известна. Такие объекты называются *неизвестными*. На выборке эталонных объектов $M \subset m$ определяется упорядоченный набор признаков $\{x_{i_1}, x_{i_2}, ..., x_{i_n}\} = X$; задаваемый экспертами предметной области. Из этого набора, если он велик, исключают признаки с малой *информативностью*, определяемые формальными способами.

Однако, следует заметить, что в большинстве теоретических и прикладных работ по распознаванию образов основное внимание уделяется разработке моделей для построения и совершенствования решающих правил распознавания. Много меньше усилий прилагается к решению проблем формирования пространств признаков, обеспечивающих оптимальные соотношения между затратами ресурсов и времени и точностью распознавания. Наиболее актуальна задача формирования таких признаковых пространств в предметных областях, где велика интенсивность потока требований на распознавание и возможны резкие колебания этой интенсивности. Назовем несколько таких областей. Это, в первую очередь медицина, микробиология, ветеринария и криминалистика. Именно в указанных областях часто возникает дефицит ресурсов и времени и, соответственно, снижается качество диагностирования.

Представляется, что проблема уменьшения затрат ресурсов и времени при построении минимальных тестов имеет очень большое значение в ряду других проблем распознавания (диагностики). Тесты используются при решении задач технической диагностики, распознавания образов, кодирования и др. [1, 2, 3]. Указанная проблема может быть разбита на две подпроблемы: а) проблему построения таких тестов для задач тестового распознавания, позволяющих уменьшить затраты ресурсов (в первую очередь, дефицитных) и времени при определении значений признаков распознаваемых объектов, т.е. в процессах распознавания [4], и б) проблему построения множеств тестов с уменьшенными затратами времени на построение тестов, что позволяет увеличить размерность используемых матриц.

Целью данной работы является разработка подходов к решению проблемы уменьшения затрат вычислительных ресурсов и времени при распознавании и диагностировании, а также построение множеств минимальных тестов или отдельных минимальных тестов на булевых матрицах. Следует обратить внимание на то, что на основе минимальных тестов могут быть выполнены построения неминимальных тупиковых тестов.

## 1. Построение минимальных тестов

Отдельные минимальные тесты можно получить, если использовать для уточнения длины минимального теста следствие из теоремы 1.

Пусть задана таблица (матрица) $Q$, содержащая значения из множества $\{0,1\}$ (булева матрица). В этой таблице строки (и столбцы) попарно различны.

<u>Определение 2</u>. [2] Множество столбцов $\{i_1, i_2, \ldots, i_k\}$ называется безусловным (диагностическим) тестом таблицы $Q$, если в подтаблице $Q(i_1, i_2, \ldots, i_k)$ нет одинаковых строк.

<u>Определение 3</u>. [2] Тест называется тупиковым, если никакое его собственное подмножество не есть тест.

<u>Определение 4</u>. Тест наименьшей длины называется минимальным. Очевидно, всякий минимальный тест – тупиковый.

Особенности предлагаемого метода построения множества минимальных тестов булевой матрицы (все элементы которой определены однозначно – неопределенных элементов нет) являются, в отличие от известных

1. Определяется длина минимального теста с помощью эвристического метода, возможно, с погрешностью, зависящей от размерности (m×n) матрицы и степенью ее разряженности. Под этой степенью здесь понимается абсолютная величина разности между количеством единиц и нулей в столбцах матрицы (всех или некоторых).

2. Производится построение тестов с вычисленной длиной минимального теста: значение этой длины (при обнаружении погрешности) корректируется, опираясь на эвристические и точные способы проверки.

Предполагается, что большие затраты на построение минимальных тестов возникают в связи с использованием точных методов при построении тестов и отсутствием проверки.

Любой тупиковый тест булевой матрицы (в том числе, минимальный) содержит хотя бы один элемент (столбец), разделяющий (различающий) каждую пару строк [1]. Существуют (могут существовать) пары строк, разделяемые только одним столбцом. Очевидно, такие столбцы обязательно должны входить в любой тест матрицы: нетупиковый, тупиковый, минимальный. Как показано в [1], целесообразно разбить множество строк матрицы на подмножества-классы с помощью обязательных элементов. В каждом классе содержатся строки, имеющие одинаковые значения обязательных элементов тестов. В разных классах все (или некоторые) обязательные элементы содержат разные значения. Такое разбиение уменьшает количество пар строк, которые должны анализироваться при построении тестов. Легко показать, что пары строк, различаемых одним обязательным элементом тестов, принадлежат множеству пар строк таких, что строки различаются по количеству единиц в них ровно на одну единицу. Предварительный анализ показал, что количество пар строк, различающихся по количеству единиц в строках ровно на единицу в неразряженных матрицах составляет ≈31%.

Для того чтобы уменьшить затраты времени при поиске обязательных элементов тестов необходимо:

а) определить количество единиц в каждой строке; б) упорядочить (отсортировать) матрицу по возрастанию значений двоичных чисел, отображаемых в строках; в) выявить пары строк, различающихся количеством единиц на одну единицу; г) выявить из пар строк, выявленных в п. «в», пары строк, различаемых (каждая) одним столбцом не менее (2-3 пары); д) произвести разбиение множества строк матрицы на подматрицы-классы; е) в сформированных подматрицах-классах произвести поиск остальных обязательных элементов; ж) если найдены дополнительные обязательные элементы в п. «е», произвести разбиение ранее созданных подматриц с помощью новых обязательных элементов.

## 2. Краткое описание метода построения множества минимальных тестов булевой матрицы

1. Сортировка строк (если тесты столбцовые) или столбцов (если тесты строковые) – по возрастанию значений двоичных чисел в строках (столбцах).
2. Поиск обязательных элементов тестов.
3. Разбиение булевой матрицы на подматрицы-классы с помощью обязательных элементов по п.2 (если они выявлены).
4. Определение длины минимальных тестов с помощью эвристической формулы на матрице и подматрицах, а также анализа пар строк в подматрицах в соответствии с теоремами.
5. Построение минимальных тестов а) выявление множеств столбцов, мощностью $t^0$, не являющихся тестами; б) определение количества минимальных тестов и составление списка минимальных тестов; в) проверка результатов и, при необходимости, корректировка.

<u>Примечание</u>. Эвристическая формула определения длины минимального теста описана автором в докладе на минской конференции в 2007 г. (с использованием некоторых корректировок). Отдельные результаты изложены автором в [6].

Уточнение длины минимального теста выполняется одним из двух указанных ниже реализаций метода. Выбирается метод с меньшим числом циклов анализа (проверки с учетом просмотра столбцов и строк), фактически, с меньшими затратами.

Число циклов *z* равно:

$$z_1 = \frac{k \cdot p(n-t_{об})!}{(t^0-1)! \cdot ((n-t_{об})-(t^0-t_{об}-1))!} = \frac{k \cdot p(n-t_{об})!}{(t^0-t_{об}-1)! \cdot (n-t^0+1)!};$$

$$z_2 = \frac{k \cdot p(n-t_{об})!}{(t^0-t_{об}-2)! \cdot (n-t^0+2)!}.$$

Теорема 1. Если в булевой матрице существует набор из $k$ столбцов, образующий хотя бы одну пару одинаковых подстрок, то такой набор из $t$ столбцов тестом матрицы не является.

Доказательство. Набор из $k$ столбцов булевой матрицы, который образует хотя бы одну пару одинаковых подстрок тестом не является, т.к. такой набор столбцов не соответствует определению теста (определение 1). □

Следствие 1. Если в булевой матрице все наборы из $k$ столбцов образуют (каждый из них) не менее одной пары одинаковых подстрок, то в этой матрице длина минимального теста $t^0 = |T^0| < k$ и, если, к тому же, существует хотя бы один тест с длиной $t = |T| = k+1$, то длина минимального теста в матрице $t^0 = |T^0| = k+1$.

Доказательство. Если выполняются условия в следствии, то в матрице в соответствии с теоремой 1, не существует ни одного теста длиной $t = |T| \leq k$ и тест T: $|T| = k+1$-есть минимальный тест. □

Теорема 2. Любые $k \geq 1$ столбцов булевой матрицы, которые образуют $p \geq 3$ одинаковых строк, не могут принадлежать одному тесту с длиной $k+1$: $|T| = k+1$, построенному из k этих столбцов и любого другого столбца матрицы.

Доказательство. Для того, чтобы в булевой матрице различить $p \geq 3$ одинаковых подстрок, необходимо, чтобы эти подстроки были попарно различимы. Число $B$ пар подстрок, в которых должны быть различимы $p$ одинаковых подстрок: $B = C_p^2 = p(p-1)/2$.

Количество пар элементов с различными значениями элементов в паре, составленных из $p$ элементов одного столбца, обеспечивается при $\psi^1 = \psi^0 = p/2$, если $p$ – четное число; при $\psi^1 - \psi^0 = 1$ (или $\psi^0 - \psi^1 = 1$), если $p$ - нечетное число, где $\psi^1(\psi^0)$ - количество единиц (нулей) в столбце. Максимальное количество пар различных по значению элементов в одном столбце, взятых из $p$ элементов: $A_{max} = \max(\psi^1 \times \psi^0) = [(p/2)^2] = [p^2/4]$, где: $[R]$ - целая часть действительного числа $R$. Минимальное количество пар одинаковых подстрок, которых останутся после добавления ($k$+1)-го столбца к набору из $k$ столбцов: $D = B - A_{max} = p(p-1)/2 - [(p/2)^2] > 0$. Так, при $p$=3: $D = 3(3-1)/2 - [(3/2)^2] = 1$; при $p$=4: $D = 4(4-1)/2 - [(4/2)^2] = 2$; при $p$=5: $D = 5(5-1)/2 - [(5/2)^2] = 4$.

Из выше изложенного можно сделать вывод: если в булевой матрице существуют $p \geq 3$ одинаковых подстрок с длиной $1 \leq k < n$ (где $n$ – число столбцов в матрице), то все эти подстроки нельзя сделать попарно различимыми с помощью одного столбца: останется хотя бы одна пара одинаковых подстрок. □

<u>Теорема 3</u>. Тупиковому тесту булевой матрицы не могут принадлежать два столбца, связанные отношением взаимно однозначного соответствия.

Доказательство. Предположим, что в тесте булевой матрицы имеются два столбца, связанные отношением взаимно однозначного соответствия. Такие столбцы содержат непременно: а) одинаковые двухэлементные подстроки матрицы, точно соответствующие одинаковым элементам в каждом из этих столбцов; б) различающиеся двухэлементные подстроки, точно соответствующие разным элементам в каждом из этих столбцов. Два таких столбца разделяют (различают) строки в одних и тех же парах строк в булевой матрице и, следовательно, один из этих столбцов является излишним для тупикового теста. □

<u>Теорема 4</u>. Для каждого столбца $x_i \in T_\tau$, принадлежащего тупиковому тесту $T_\tau \subset X$ булевой матрицы, существует, хотя бы одна пара строк $(s_{j_1}, s_{j_2})$: $s_{j_1}, s_{j_2} \in S$, различаемых элементами столбца $x_i$ и никакого другого столбца $x_v \in T_\tau \setminus x_i$, принадлежащего этому же тупиковому тесту $T_\tau$:

$$(\forall T_\tau \subset X)(\forall x_i \in T_\tau) \to \exists \left((s_{j_1}, s_{j_2}) \subset S\right)\left(\left(a_{s_{j_1}x_i} \neq a_{s_{j_2}x_i}\right) \& (a_{s_{j_1}x_v} = a_{s_{j_2}x_v})\right).$$

Доказательство. Предположим, что утверждение, сформулированное в теореме ложно и верно следующее утверждение:

$$\neg\left((\forall T_\tau \subset X)(\forall x_i \in T_\tau) \to \exists \left((s_{j_1}, s_{j_2}) \subset S\right)\left(\left(a_{s_{j_1}x_i} \neq a_{s_{j_2}x_i}\right) \& (a_{s_{j_1}x_v} = a_{s_{j_2}x_v})\right)\right).$$

Выполнив преобразования этого утверждения, получим:

$$(\forall T_\tau \subset X)(\forall x_i \in T_\tau) \& (\forall (s_{j_1}, s_{j_2}) \subset S)\left(\left(a_{s_{j_1}x_i} = a_{s_{j_2}x_i}\right) \& (a_{s_{j_1}x_v} \neq a_{s_{j_2}x_v})\right) \vee \left(\left(a_{s_{j_1}x_i} \neq a_{s_{j_2}x_i}\right) \& (a_{s_{j_1}x_v} \neq a_{s_{j_2}x_v})\right)$$

Полученное утверждение противоречит определениям тупикового теста и теста булевой матрицы. Следовательно, утверждение, сформулированное в данной теореме истинно. □

<u>Теорема 5</u>. Тест $T$ булевой матрицы, имеющей длину $t^0$ является минимальным тестом тогда и только тогда, когда все тесты с длиной $t^0$ данной матрицы являются тупиковыми.

$$(\forall T \subset X)\left((|T| = |T_\tau| = t^0) \Leftrightarrow (T_\tau = T_{min})\right), \qquad (1)$$

Где $X$ – множество столбцов матрицы $Q$; $T_\tau$ - тупиковый тест; $T_{min}$ - минимальный тест;
$t^0$ - длина минимального теста.

Доказательство. 1) Достаточность. Пусть условие в левой части утверждения (1) выполняется, а условие в правой части – не выполняется. Возможны следующие варианты:
а) $|T_\tau| < |T_{min}|$, что противоречит определению 3.
б) $|T_\tau| > |T_{min}|$ – противоречит левой части (1) в теореме. Следовательно, утверждение в теореме верно.
2 Необходимость. Пусть условие в левой части утверждения (1) не выполняется то есть существует хотя бы один тест : $(T' \subset X): |T'| = t^0$, не являющийся тупиковым, и существует, следовательно, тупиковый тест $T'_\tau \subset T'$; $|T'_\tau| < t^0$, что противоречит утверждению теоремы. □

### 3. Примеры построения множества минимальных тестов
**3.1. Пример 1**. **Матрица** $Q$: $m=25$; $n=10$. Строки матрицы - 10-разрядные числа, полученные с помощью генератора псевдослучайных чисел

| 1 | 2 | 3 | 4 | 5 | 6 | 7 | 8 | 9 | 10 | S | кед |
|---|---|---|---|---|---|---|---|---|---|---|---|
| 1 | 1 | 0 | 1 | 1 | 0 | 0 | 0 | 0 | 0 | 1 | 4 |
| 0 | 0 | 0 | 1 | 0 | 0 | 0 | 0 | 0 | 1 | 2 | 2 |
| 1 | 1 | 1 | 1 | 0 | 0 | 1 | 1 | 1 | 0 | 3 | 7 |
| 0 | 0 | 0 | 0 | 1 | 1 | 0 | 1 | 0 | 1 | 4 | 4 |
| 0 | 0 | 0 | 1 | 1 | 0 | 0 | 1 | 0 | 1 | 5 | 4 |
| 1 | 0 | 0 | 1 | 1 | 1 | 0 | 0 | 0 | 1 | 6 | 5 |
| 1 | 0 | 1 | 0 | 0 | 0 | 1 | 0 | 0 | 1 | 7 | 4 |
| 1 | 1 | 0 | 0 | 1 | 0 | 0 | 0 | 1 | 0 | 8 | 4 |
| 1 | 0 | 1 | 1 | 0 | 1 | 0 | 1 | 1 | 0 | 9 | 6 |
| 0 | 0 | 0 | 1 | 1 | 1 | 1 | 0 | 1 | 1 | 10 | 6 |
| 0 | 0 | 0 | 0 | 0 | 1 | 0 | 0 | 0 | 0 | 11 | 1 |
| 0 | 0 | 0 | 0 | 0 | 0 | 1 | 1 | 1 | 0 | 12 | 3 |
| 1 | 0 | 1 | 0 | 0 | 1 | 0 | 1 | 0 | 0 | 13 | 4 |
| 0 | 0 | 1 | 0 | 1 | 0 | 1 | 0 | 0 | 1 | 14 | 4 |
| 1 | 1 | 1 | 0 | 0 | 1 | 1 | 0 | 1 | 0 | 15 | 6 |
| 0 | 0 | 0 | 1 | 0 | 0 | 1 | 1 | 0 | 0 | 16 | 3 |
| 1 | 0 | 0 | 1 | 1 | 0 | 1 | 1 | 1 | 1 | 17 | 7 |
| 0 | 1 | 0 | 0 | 0 | 0 | 1 | 0 | 1 | 0 | 18 | 3 |
| 0 | 0 | 0 | 0 | 0 | 0 | 1 | 0 | 1 | 1 | 19 | 3 |
| 0 | 1 | 1 | 0 | 0 | 0 | 1 | 0 | 1 | 1 | 20 | 5 |
| 1 | 0 | 0 | 1 | 0 | 1 | 0 | 0 | 0 | 1 | 21 | 4 |
| 0 | 0 | 0 | 0 | 0 | 0 | 1 | 1 | 1 | 1 | 22 | 4 |
| 0 | 1 | 0 | 0 | 1 | 0 | 0 | 0 | 0 | 1 | 23 | 3 |
| 0 | 1 | 1 | 0 | 1 | 1 | 0 | 1 | 1 | 1 | 24 | 7 |
| 0 | 0 | 0 | 1 | 1 | 0 | 0 | 0 | 0 | 1 | 25 | 3 |

Строки $S_j$, которые содержат $r$-единиц
$r=1$: $S_{11}$; $r=2$: $S_2$; $r=3$: $S_{12}, S_{16}, S_{18}, S_{19}, S_{23}, S_{25}$; $r=4$: $S_1, S_4, S_5, S_7, S_8, S_{13}, S_{14}, S_{21}$; $r=5$: $S_6, S_{20}$, $r=6$: $S_9, S_{10} S_{15}$, $r=7$: $S_3, S_{17}, S_{24}$.

Общее количество пар : $m(m-1)/2 = 25 \cdot 24/2 = 300$. Доля просматриваемых пар от их общего количества $94/300 = 0,313333 = 31\%$.

Пары строк, которые различаются одним столбцом и соответствующие им столбцы: $((S_2 - S_{25}), (S_6 - S_{21})) \to x_5$;    $((S_5 - S_{25}), (S_{19} - S_{22})) \to x_8$;
$(S_{12} - S_{22}) \to x_{10}$.

Булева матрица $Q_{25,10}$, упорядоченная по возрастанию значений двоичных чисел в строках.

| 1 | 2 | 3 | 4 | 5 | 6 | 7 | 8 | 9 | 10 | S | кед |
|---|---|---|---|---|---|---|---|---|---|---|---|
| 0 | 0 | 0 | 0 | 0 | 0 | 1 | 0 | 1 | 1 | 19 | 3 |
| 0 | 0 | 0 | 0 | 0 | 0 | 1 | 1 | 1 | 0 | 12 | 3 |
| 0 | 0 | 0 | 0 | 0 | 0 | 1 | 1 | 1 | 1 | 22 | 4 |
| 0 | 0 | 0 | 0 | 0 | 1 | 0 | 0 | 0 | 0 | 11 | 1 |
| 0 | 0 | 0 | 0 | 1 | 1 | 0 | 1 | 0 | 1 | 4 | 4 |
| 0 | 0 | 0 | 1 | 0 | 0 | 0 | 0 | 0 | 1 | 2 | 2 |
| 0 | 0 | 0 | 1 | 0 | 0 | 1 | 1 | 0 | 0 | 16 | 3 |
| 0 | 0 | 0 | 1 | 1 | 0 | 0 | 0 | 0 | 1 | 25 | 3 |
| 0 | 0 | 0 | 1 | 1 | 0 | 0 | 1 | 0 | 1 | 5 | 4 |
| 0 | 0 | 0 | 1 | 1 | 1 | 1 | 0 | 1 | 1 | 10 | 6 |
| 0 | 0 | 1 | 0 | 1 | 0 | 1 | 0 | 0 | 1 | 14 | 4 |
| 0 | 1 | 0 | 0 | 0 | 0 | 1 | 0 | 1 | 0 | 18 | 3 |
| 0 | 1 | 0 | 0 | 1 | 0 | 0 | 0 | 0 | 1 | 23 | 3 |
| 0 | 1 | 1 | 0 | 0 | 0 | 1 | 0 | 1 | 1 | 20 | 5 |
| 0 | 1 | 1 | 0 | 1 | 1 | 0 | 1 | 1 | 1 | 24 | 7 |
| 1 | 0 | 0 | 1 | 0 | 1 | 0 | 0 | 0 | 1 | 21 | 4 |
| 1 | 0 | 0 | 1 | 1 | 0 | 1 | 1 | 1 | 1 | 17 | 7 |
| 1 | 0 | 0 | 1 | 1 | 1 | 0 | 0 | 0 | 1 | 6 | 5 |
| 1 | 0 | 1 | 0 | 0 | 0 | 1 | 0 | 0 | 1 | 7 | 4 |
| 1 | 0 | 1 | 0 | 0 | 1 | 0 | 1 | 0 | 0 | 13 | 4 |
| 1 | 0 | 1 | 1 | 0 | 1 | 0 | 1 | 1 | 0 | 9 | 6 |
| 1 | 1 | 0 | 0 | 1 | 0 | 0 | 0 | 1 | 0 | 8 | 4 |
| 1 | 1 | 0 | 1 | 1 | 0 | 0 | 0 | 0 | 0 | 1 | 4 |
| 1 | 1 | 1 | 0 | 0 | 1 | 1 | 0 | 1 | 0 | 15 | 6 |
| 1 | 1 | 1 | 1 | 0 | 0 | 1 | 1 | 1 | 0 | 3 | 7 |

В результате анализа строк, различающихся числом единиц в первых 9 строках упорядоченной матрицы, выявлены 3 одноэлементных теста в 16 парах строк: 19-22; 12-22; 19-4; 12-4; 19-5; 12-5; 16-22; 25-22; 16-4; 25-4; 16-5; 25-5; 19-2; 22-2; 16-2; 25-2.

Уменьшения затрат при поиске остальных единичных тестов можно добиться, если дальнейший поиск вести в подматрицах матрицы $Q$, полученных разбиением матрицы с помощью найденных одноэлементных тестов для пар строк.

Получены следующие подматрицы

$(x_5, x_8, x_{10})$

$\underline{0\ \ 0\ \ 0}$

$$Q_1 = \begin{array}{c} \begin{array}{ccccccc} 1 & 2 & 3 & 4 & 6 & 7 & 9 \end{array} \ \ \ s \ \ \ \text{кед} \\ \begin{bmatrix} 0 & 0 & 0 & 0 & 1 & 0 & 0 \\ 1 & 1 & 1 & 0 & 1 & 1 & 1 \\ 0 & 1 & 0 & 0 & 0 & 1 & 1 \end{bmatrix} \begin{array}{c} 11 \\ 15 \\ 18 \end{array} \begin{array}{c} 1 \\ 6 \\ 3 \end{array} \end{array}$$

$\underline{0\ \ 0\ \ 1}$

$$Q_2 = \begin{bmatrix} 0 & 0 & 0 & 1 & 0 & 0 & 0 \\ 1 & 0 & 1 & 0 & 0 & 1 & 0 \\ 0 & 0 & 0 & 0 & 0 & 1 & 1 \\ 0 & 1 & 1 & 0 & 0 & 1 & 1 \\ 1 & 0 & 0 & 1 & 1 & 0 & 0 \end{bmatrix} \begin{array}{c} 2 \\ 7 \\ 19 \\ 20 \\ 21 \end{array} \begin{array}{c} 1 \\ 3 \\ 2 \\ 4 \\ 3 \end{array}$$

$\underline{0\ \ 1\ \ 0}$

$$Q_3 = \begin{bmatrix} 1 & 1 & 1 & 1 & 0 & 1 & 1 \\ 1 & 0 & 1 & 1 & 1 & 0 & 1 \\ 0 & 0 & 0 & 0 & 0 & 1 & 1 \\ 1 & 0 & 1 & 0 & 1 & 0 & 0 \\ 0 & 0 & 0 & 1 & 0 & 1 & 0 \end{bmatrix} \begin{array}{c} 3 \\ 9 \\ 12 \\ 13 \\ 16 \end{array} \begin{array}{c} 6 \\ 5 \\ 2 \\ 3 \\ 2 \end{array}$$

$\underline{1\ \ 0\ \ 0}$

$$Q_5 = \begin{bmatrix} 1 & 1 & 0 & 1 & 0 & 0 & 0 \\ 1 & 1 & 0 & 0 & 0 & 0 & 1 \end{bmatrix} \begin{array}{c} 1 \\ 8 \end{array} \begin{array}{c} 3 \\ 3 \end{array}$$

$\underline{1\ \ 0\ \ 1}$

$$Q_6 = \begin{bmatrix} 1 & 0 & 0 & 1 & 1 & 0 & 0 \\ 0 & 0 & 0 & 1 & 1 & 1 & 1 \\ 0 & 0 & 1 & 0 & 0 & 1 & 0 \\ 0 & 1 & 0 & 0 & 0 & 0 & 0 \\ 0 & 0 & 0 & 1 & 0 & 0 & 0 \end{bmatrix} \begin{array}{c} 6 \\ 10 \\ 14 \\ 23 \\ 25 \end{array} \begin{array}{c} 3 \\ 4 \\ 1 \\ 1 \\ 1 \end{array}$$

$\underline{1\ \ 1\ \ 1}$

$$Q_7 = \begin{bmatrix} 0 & 0 & 0 & 0 & 1 & 0 & 0 \\ 0 & 0 & 0 & 1 & 0 & 0 & 0 \\ 1 & 0 & 0 & 1 & 0 & 1 & 1 \\ 0 & 1 & 1 & 0 & 1 & 0 & 1 \end{bmatrix} \begin{array}{c} 4 \\ 5 \\ 17 \\ 24 \end{array} \begin{array}{c} 1 \\ 1 \\ 4 \\ 4 \end{array}$$

Анализ 8 пар строк в подматрицах: $Q_2$: $(7-19,\ 19-21,\ 7-20,\ 20-21)$, $Q_3$: $(3-9, 12-13,\ 13-16)$, $Q_6$: $(6-10)$ не выявил дополнительно ни одного теста (одноэлементного). Всего проанализированы 24 пары строк, т.е. в $94/24 \approx 4$ раза.

### 3.1.1. Определение длины минимального теста матрицы $Q_{25,10}$ с помощью эвристического алгоритма $m=25$, $\widehat{m}=300$.

Количество единиц и нулей в строках матрицы $Q$:
$\psi_1^1 = \psi_8^1 = 10;$   $\psi_1^0 = \psi_8^0 = 25 - 10 = 15;$   $\psi_2^1 = \psi_3^1 = 8;$
$\psi_2^0 = \psi_3^0 = 25 - 8 = 17;$   $\psi_4^1 = \psi_5^1 = 11;$   $\psi_4^0 = \psi_5^0 = 25 - 11 = 14;$
$\psi_6^1 = 9;$ $\psi_6^0 = 25 - 9 = 16;$   $\psi_7^1 = \psi_9^1 = 12;$   $\psi_7^0 = \psi_9^0 = 25 - 12 = 13;$
$\psi_{10}^1 = 15;$   $\psi_{10}^0 = 25 - 15 = 10.$

Количество нулей в столбцах матрицы различий для матрицы Q (построение в данном методе не выполняется)
$\widehat{\psi}_1^0 = \widehat{\psi}_8^0 = \widehat{\psi}_{10}^0 = 300 - 15 \cdot 1 = 150;$
$\widehat{\psi}_2^0 = \widehat{\psi}_3^0 = 300 - 8 \cdot 17 = 164;$   $\widehat{\psi}_4^0 = \widehat{\psi}_5^0 = 300 - 11 \cdot 14 = 146;$
$\widehat{\psi}_6^0 = 300 - 9 \cdot 16 = 156;$   $\widehat{\psi}_7^0 = \widehat{\psi}_9^0 = 300 - 12 \cdot 13 = 144;$

Составляем список столбцов, упорядоченный по возрастанию значений $\widehat{\psi}_i^0$:

$x_7, x_9, x_4, x_5, x_1, x_8, x_{10}, x_6, x_2, x_3.$

$\beta_{t^0} = \frac{144^2 \cdot 146^2 \cdot 150^2}{300^7} > \frac{1}{300} > \beta_{t^0+1} = \beta_{t^0} \cdot \psi_{1\text{ср}}^0 > \beta_{t^0+1} = \frac{\beta_{t^0}}{300};$

$\beta_{t^0} = 0{,}0068211 > 0{,}0333333 > \frac{0{,}0032741 \cdot 144}{300} > 0{,}0032741 = 0{,}0032741.$

$t^0 = 7$ или $\beta_{t^0} = \left(\frac{144}{300}\right)^7 = 530839 > 0{,}00333 > 530839 \cdot \frac{144}{300} = 254802.$

Определение длины локального минимального теста ($t_{\text{лок}}^0 = t^0 - r$), где $r$ – количество обязательных элементов в тесте на подматрице $Q'$

$Q'_{10,7} = Q_2 \cup Q_3$,  $m=10$; $\widehat{m} = 45$; $\widehat{m}^{-1} = 0{,}0222222.$
$\psi_1^1 = \psi_3^1 = \psi_4^1 = \psi_9^1 = 5;$   $\psi_2^1 = 2;$   $\psi_6^1 = 3;$   $\psi_7^1 = 6;$
$\psi_1^0 = \psi_3^0 = \psi_4^0 = \psi_9^0 = 5;$   $\psi_2^0 = 8;$   $\psi_6^0 = 7;$   $\psi_7^0 = 4;$
$\widehat{\psi}_1^1 = \widehat{\psi}_3^1 = \widehat{\psi}_4^1 = \widehat{\psi}_9^1 = 25;$   $\widehat{\psi}_2^1 = 16;$   $\widehat{\psi}_6^1 = 21;$   $\widehat{\psi}_7^1 = 24;$
$\widehat{\psi}_1^0 = \widehat{\psi}_3^0 = \widehat{\psi}_4^0 = \widehat{\psi}_9^0 = 20;$   $\widehat{\psi}_2^0 = 29;$   $\widehat{\psi}_6^0 = 24;$   $\widehat{\psi}_7^0 = 21;$
$\beta_{t^0} = \left(\frac{20}{45}\right)^4 > 0{,}00222222 > \left(\frac{20}{45}\right)^4 \cdot \frac{20}{45} = \beta_{t^0+1}.$
$0{,}039 > 0{,}00222222 > 0{,}0173415 = \beta_{t^0+1}$

Или: $\beta_{t^0} = \left(\frac{20}{45}\right)^4 > 0{,}00222222 > \left(\frac{20}{45}\right)^4 \cdot \frac{21}{45} = \beta_{t^0+1}$
$0{,}039 > 0{,}00222222 > 0{,}0182085 = \beta_{t^0+1}.$

### 3.1.2. Уточнение длины локального минимального теста: $t^0 = |T^0|$

Наборы из $k$ столбцов, образующие $p$ одинаковых подстрок матрицы $Q_{25,10}$,  $k=3$, $p=2$. Метод опирается на следствие из теоремы 1, приведенной выше в работе.

123 - $Q_2(19,2)$         267 - $Q_2$
124 - $Q_6(25,10)$        269 - $Q_2$
126 - $Q_2(19,2)$         279 - $Q_2$
127 - $Q_3$               346 - $Q_6$

129 - $Q_6$         347 - $Q_6$
134 - $Q_3$         349 - $Q_6$
136 - $Q_3$         367 - $Q_6$
137 - $Q_3$         369 - $Q_6$
139 - $Q_3$         379 - $Q_6$
146 - $Q_2$         467 - $Q_2$
147 - $Q_2$         469 - $Q_2$
149 - $Q_2$         479 - $Q_2$
167 - $Q_2$         679 - $Q_2$
169 - $Q_2$
179 - $Q_2$
234 - $Q_2$
236 - $Q_2$
237 - $Q_2$
239 - $Q_2$
246 - $Q_2$
247 - $Q_2$
249 - $Q_2$

Контрольная проверка: $C_7^3 = \frac{4 \cdot 5 \cdot 6 \cdot 7}{6 \cdot 4} = 35$.

Вывод: $t^0 > 3$

Список троек столбцов составлен для визуального контроля, учитывающего пары одинаковых подстрок; дублирующие в данных список не включены. [Соловьев Н.А., стр. 72].

### 3.1.3. Наборы из $k$ столбцов, образующие $p$ одинаковых подстрок матрицы $Q_{25,10}$, $k$=2, $p$=3.

Метод опирается на теорему 2, приведенной выше в работе. (Пары столбцов, образующие тройки одинаковых подстрок в подматрицах матрицы $Q_{25,10}$ для уточнения числа ($t^0 - 1$)).

$(1,2) - Q_2 \to s: (10, 14, 25)$
$(1,3) - Q_3 \to s: (3, 9, 13)$
$(1,6) - Q_2 \to s: (2, 19, 20)$
$(2,3) - Q_6 \to s: (6, 10, 25)$
$(2,6) - Q_2 \to s: (2, 7, 19)$
$(2,9) - Q_2 \to s: (2, 7, 21)$
$(4,6) - Q_2 \to s: (7, 19, 20)$
$(4,7) - Q_2 \to s: (7, 19, 20)$
$(6,7) - Q_2 \to s: (7, 19, 20)$

<u>Булева матрица:</u> $Q$: $m$=25; $n$=10
Обязательные элементы тестов: $\{x_5, x_8, x_{10},\}$
$\hat{\psi}_1^1 = \hat{\psi}_8^1 = \hat{\psi}_{10}^1 = 150$; $\hat{\psi}_2^1 = \hat{\psi}_3^1 = 136$; $\hat{\psi}_4^1 = \hat{\psi}_5^1 = 154$;
$\hat{\psi}_6^1 = 144$; $\hat{\psi}_7^1 = \hat{\psi}_9^1 = 156$.

Четверки, не принадлежащие минимальным тестам

<u>1234</u> - $Q_6$: $(s_{10} - s_{25})$; <u>1236</u> - $Q_2$: $(s_2 - s_{19})$; $Q_3$: $(s_{12} - s_{16})$, $(s_9 - s_3)$; $Q_5$: $(s_1 - s_8)$; <u>1237</u> - $Q_3$: $(s_9 - s_{13})$; $(s_{12} - s_{16})$; $Q_5$: $(s_1 - s_8)$; $Q_7$: $(s_9 - s_{13})$; <u>1239</u> - $Q_7$: $(s_4 - s_5)$; <u>1267</u> - $Q_3$: $(s_9 - s_{13})$; $(s_{12} - s_{16})$; $Q_5$: $(s_1 - s_8)$; <u>1349</u> - $Q_3$: $(s_3 - s_9)$;
<u>1269</u> - $Q_6$: $(s_{14} - s_{25})$; <u>1279</u> - $Q_7$: $(s_4 - s_5)$; <u>1367</u> - $Q_3$: $(s_9 - s_{13})$; $(s_{12} - s_{16})$; $Q_5$: $(s_1 - s_8)$; $Q_6$: $(s_{23} - s_{25})$; <u>1369</u> - $Q_6$: $(s_{23} - s_{25})$;
<u>1379</u> - $Q_6$: $(s_{23} - s_{25})$; $Q_7$: $(s_4 - s_5)$; <u>1467</u> - $Q_7$: $(s_4 - s_{24})$; $Q_2$: $(s_{19} - s_{20})$;
<u>1469</u> - $Q_2$: $(s_{19} - s_{20})$; $Q_6$: $(s_{14} - s_{23})$; <u>1479</u> - $Q_2$: $(s_{19} - s_{20})$;
<u>1679</u> - $Q_2$: $(s_{19} - s_{20})$; $Q_6$: $(s_{23} - s_{25})$;
<u>2346</u> - $Q_6$: $(s_6 - s_{10})$; $Q_7$: $(s_5 - s_{17})$; <u>2347</u> - $Q_2$: $(s_2 - s_{21})$; $Q_6$: $(s_6 - s_{25})$;
<u>2349</u> - $Q_2$: $(s_2 - s_{21})$; $Q_6$: $(s_6 - s_{25})$; <u>2367</u> - $Q_3$: $(s_9 - s_{13})$; $(s_{12} - s_{16})$; $Q_5$: $(s_1 - s_8)$; <u>2379</u> - $Q_2$: $(s_2 - s_{21})$; $Q_6$: $(s_6 - s_{25})$; $Q_7$: $(s_4 - s_5)$;
<u>2467</u> - $Q_2$: $(s_7 - s_{19})$; <u>2479</u> - $Q_1$: $(s_{15} - s_{18})$; $Q_2$: $(s_2 - s_{21})$; $Q_6$: $(s_6 - s_{25})$;
<u>3467</u> - $Q_2$: $(s_7 - s_{20})$; <u>3479</u> - $Q_2$: $(s_2 - s_{21})$; $Q_6$: $(s_6 - s_{25})$;
<u>3679</u> – $Q_6$: $(s_{23} - s_{25})$; <u>4679</u> – $Q_2$: $(s_{19} - s_{20})$.

В результате обнаружено количество четверок – столбцов – не тестов 26. Всего четверок столбцов в матрице: $7!/(4! \cdot 3!) = 35$. Количество четверок столбцов тестов: 35-26=9.

Четверки столбцов, принадлежащие минимальному тесту:
(1246); (1247); (1249); (1346); (1347); (2369); (2469); (2679); (3469).

### 3.1.4. Проверка тестов на тупиковость

<u>1246</u>: $\underline{x_1} - Q_2$: $(s_7 - s_{19})$; $Q_3$: $(s_6 - s_{10})$; $Q_7$: $(s_5 - s_{17})$;
$\underline{x_2} - Q_2$: $(s_{19} - s_{20})$; $Q_6$: $(s_{14} - s_{23})$; $Q_7$: $(s_4 - s_{24})$; $\underline{x_4} - Q_2$: $(s_2 - s_{19})$; $Q_3$: $(s_9 - s_{13}$; $Q_5$:$s_1 - s_8$; $Q_6$:$s_{14} - s_{25}$;
$\underline{x_6} - Q_6$: $(s_{10} - s_{25})$;
<u>1247</u>: $\underline{x_1} - Q_1$: $(s_{15} - s_{18})$; $Q_2$: $(s_2 - s_{21})$, $(s_7 - s_{19})$; $Q_6$: $(s_6 - s_{25})$;
$\underline{x_2} - Q_2$: $(s_{19} - s_{20})$; $Q_7$: $(s_4 - s_{24})$;
$\underline{x_4} - Q_3$: $(s_9 - s_{13})$, $(s_{12} - s_{16})$; $Q_5$: $(s_1 - s_8)$; $Q_6$: $(s_{10} - s_{14})$; $Q_7$: $(s_4 - s_5)$;
$\underline{x_7} - Q_6$: $(s_{10} - s_{25})$;
<u>1249</u>: $\underline{x_1} - Q_1$: $(s_{15} - s_{18})$; $Q_2$: $(s_2 - s_{21})$; $Q_6$: $(s_6 - s_{25})$;
$\underline{x_2} - Q_2$: $(s_{19} - s_{20})$; $Q_3$: $(s_3 - s_9)$; $Q_6$: $(s_{14} - s_{23})$;
$\underline{x_4} - Q_6$: $(s_{14} - s_{25})$; $Q_7$: $(s_4 - s_5)$;
$\underline{x_9} - Q_6$: $(s_{10} - s_{25})$;
<u>1346</u>: $\underline{x_1} - Q_2$: $(s_7 - s_{20})$; $Q_6$: $(s_6 - s_{10})$; $Q_7$: $(s_5 - s_{17})$;
$\underline{x_3} - Q_2$: $(s_{19} - s_{20})$; $Q_6$: $(s_{14} - s_{23})$; $Q_7$: $(s_4 - s_{24})$;
$\underline{x_4} - Q_2$: $(s_2 - s_{19})$; $Q_3$: $(s_9 - s_{13})$, $(s_{12} - s_{16})$; $Q_5$: $(s_1 - s_8)$; $Q_6$: $(s_{23} - s_{25})$;
$\underline{x_6} - Q_3$: $(s_3 - s_9)$; $Q_6$: $(s_{10} - s_{25})$; $Q_1$: $(s_{11} - s_{15})$;
<u>1347</u>: $\underline{x_1} - Q_2$: $(s_2 - s_{21})$, $(s_7 - s_{20})$; $Q_6$: $(s_6 - s_{25})$;

$\underline{x_3} - Q_2: (s_{19} - s_{20}); Q_7: (s_4 - s_{24});$
$\underline{x_4} - Q_3: (s_9 - s_{13}), (s_{12} - s_{16}); Q_5: (s_1 - s_8);$
$\underline{x_7} - Q_1: (s_{11} - s_{18}); Q_3: (s_3 - s_9);$
$\underline{2369}:\ \underline{x_2} - Q_6: (s_{23} - s_{25});$
$\underline{x_3} - Q_2: (s_2 - s_7);$
$\underline{x_6} - Q_2: (s_2 - s_{21}); Q_7: (s_4 - s_5);$
$\underline{x_9} - Q_2: (s_2 - s_{19}); Q_3: (s_9 - s_{13});\ Q_5: (s_1 - s_8); Q_6: (s_6 - s_{10}); Q_7: (s_5 - s_{17});$
$\underline{2469}:\ \underline{x_2} - Q_2: (s_{19} - s_{20}); Q_6: (s_{14} - s_{23});$
$\underline{x_4} - Q_2: (s_2 - s_7); Q_6: (s_{14} - s_{25});$
$\underline{x_6} - Q_1: (s_{15} - s_{18}); Q_2: (s_2 - s_{21}); Q_6: (s_6 - s_{25});$
$\underline{x_9} - Q_2: (s_7 - s_{19}); Q_6: (s_6 - s_{10});\ Q_7: (s_5 - s_{17});$
$\underline{2679}:\ \underline{x_2} - Q_2: (s_{19} - s_{20}); Q_3: (s_3 - s_{12}); Q_6: (s_{23} - s_{25});$
$\underline{x_6} - Q_1: (s_{15} - s_{18}); Q_2: (s_2 - s_{21}); Q_6: (s_6 - s_{25}); Q_7: (s_4 - s_5);$
$\underline{x_7} - Q_2: (s_2 - s_7); Q_6: (s_{14} - s_{25});$
$\underline{x_9} - Q_2: (s_7 - s_{19}); Q_3: (s_9 - s_{13}), (s_{12} - s_{16});\ Q_5: (s_1 - s_8);$
$\underline{369}:\ \underline{x_3} - Q_2: (s_{19} - s_{20}); Q_6: (s_{14} - s_{23});$
$\underline{x_4} - Q_6: (s_{23} - s_{25});$
$\underline{x_6} - Q_2: (s_2 - s_{21}); Q_3: (s_3 - s_9);\ Q_6: (s_6 - s_{25});$
$\underline{x_9} - Q_2: (s_7 - s_{20}); Q_6: (s_6 - s_{10});$

### 3.2. Пример 2. Построение минимальных тестов для $M_{50,10}$

Разбиение строк $s_{26}, \ldots, s_{75}$ матрицы $M_{160,10}$ по столбцам: $x_2, x_3, x_4, x_6, x_7, x_{10}$, являющимся обязательными элементами тестов матрицы M на подматрицы $M_1, M_2, M_3, M_4, M_5, M_6, M_7, M_8$; не приведены однострочные подматрицы, полученные по результатам разбиения.

Длина минимальных тестов, вычисленная по эвристической формуле - а) $t^0 = 10$; б) $t^0 = 9$. Длина минимальных локальных тестов в соответствии с приведенными теоремами равна 2. Длина интегральных минимальных тестов матрицы $M_{50,10}$: $t_и^0 = 6 + 2 = 8$.

101101
$$M_1 = \begin{bmatrix} 1 & 5 & 8 & 9 & s \\ 1 & 1 & 1 & 1 \\ 1 & 1 & 0 & 0 \end{bmatrix} \begin{matrix} 33 \\ 55 \end{matrix}$$

101110
$$M_2 = \begin{bmatrix} 1 & 1 & 0 & 0 \\ 1 & 0 & 0 & 1 \end{bmatrix} \begin{matrix} 44 \\ 47 \end{matrix}$$

101111
$$M_3 = \begin{bmatrix} 1 & 1 & 1 & 0 \\ 0 & 1 & 0 & 0 \end{bmatrix} \begin{matrix} 35 \\ 60 \end{matrix}$$

001100
$$M_4 = \begin{bmatrix} 1 & 1 & 1 & 1 \\ 1 & 0 & 0 & 0 \end{bmatrix} \begin{matrix} 29 \\ 66 \end{matrix}$$

001001
$$M_5 = \begin{bmatrix} 1 & 1 & 0 & 1 \\ 1 & 0 & 0 & 0 \end{bmatrix} \begin{matrix} 34 \\ 36 \end{matrix}$$

100001
$$M_6 = \begin{bmatrix} 1 & 0 & 0 & 1 \\ 1 & 1 & 1 & 0 \end{bmatrix} \begin{matrix} 28 \\ 39 \end{matrix}$$

110001
$$M_7 = \begin{bmatrix} 1 & 0 & 1 & 1 \\ 0 & 1 & 1 & 0 \end{bmatrix} \begin{matrix} 52 \\ 62 \end{matrix}$$

001010
$$M_8 = \begin{bmatrix} 1 & 1 & 0 & 1 \\ 0 & 1 & 1 & 1 \end{bmatrix} \begin{matrix} 38 \\ 63 \end{matrix}$$

Пары столбцов в подматрицах $M_1, \ldots, M_8$, не являющиеся минимальными тестами $(x_1, x_8), x_5, x_9$.

Количество минимальных локальных тестов: $C_4^2 - 2 = \frac{3 \cdot 4}{2} - 2 = 6 - 2 = 4$.

(Количество сочетаний двух столбцов из четырех: $C_4^2 = 6$. Количество пустых пар – 2, количество локальных минимальных тестов: 6-2=4).

Локальные минимальные тесты: $(x_1, x_5), (x_1, x_9), (x_5, x_8), (x_8, x_9)$.

Интегральные минимальные тесты: $(x_1, x_2, x_3, x_4, x_5, x_6, x_7, x_{10})$; $(x_1, x_2, x_3, x_4, x_6, x_7, x_9, x_{10})$; $(x_2, x_3, x_4, x_5, x_6, x_7, x_8, x_{10})$; $(x_2, x_3, x_4, x_6, x_7, x_8, x_9, x_{10})$.

Число пар строк в матрице $M$: $\widehat{m} = 50 \times 49/2 = 1225$.

Число пар строк в подматрицах $M_1,\ldots,M_8$ равно: $\sum \hat{m} = 8$, то есть в 153 раза меньше.

## Заключение

Работа посвящена задаче уменьшения затрат ресурсов и времени при распознавании и диагностике путем разработки метода построения множества минимальных тестов или отдельных минимальных тестов а также минимизации затрат на их формирование.

Описаны задачи распознавания и диагностики, особенности таких задач.

Изложены исследования по проблеме и достигнутые практические результаты.

Предложен новый метод построения множеств минимальных тестов или отдельных минимальных тестов, обеспечивающий качество тестов при существенно меньших затратах вычислительных ресурсов.

Приведены примеры (2 примера), показывающие реализацию предложенного метода на матрицах различной размерности; также даны уточнение длины локального минимального теста, и проверка теста на тупиковость.

Представляется, что необходимо провести испытание метода на потоке случайных булевых матриц различной размерности с целью оценки эффективности предложенного метода.

Статья посвящена проблеме уменьшения затрат вычислительных ресурсов и времени в задачах построения минимальных тестов булевых матриц. На основе минимальных тестов могут быть выполнены построения неминимальных тупиковых тестов.